# Faster Algorithm of String Comparison

**Qi Xiao Yang**
Institute of High Performance of Computing
89B Science Park Drive#01-05/08 the Rutherford
Singapore 118261
qixy@ihpc.nus.edu.sg or cwl1012@hotmail.com
tel: (65)7709265

**Sung Sam Yuan, Li Zhao,Lu Chun and Sun Peng**
School of Computing
National University of Singapore
3 Science Drive 2, Singapore 117543
{ssung,lizhao,luchun,sunpeng1}@comp.nus.edu.sg
tel: (65)8746148

## Abstract

In many applications, it is necessary to determine the string similarity*. Text comparison now appears in many disciplines such as compression, pattern recognition, computational biology, Web searching and data cleaning. Edit distance[WF74] approach is a classic method to determine Field Similarity. A well known dynamic programming algorithm [GUS97] is used to calculate edit distance with the time complexity O(nm). (for worst case, average case and even best case) Instead of continuing with improving the edit distance approach, [LL+99] adopted a brand new approach---token-based approach. Its new concept of token-base-----retain the original semantic information, good time complex----O(nm) (for worst, average and best case) and good experimental performance make it a milestone paper in this area. Further study indicates that there is still room for improvement of its Field Similarity algorithm. Our paper is to introduce a package of substring-based new algorithms to determine Field Similarity. Combined together, our new algorithms not only achieve higher accuracy but also gain the time complexity O(knm) (k<0.75) for worst case, O($\beta$ *n) where $\beta$ <6 for average case and O(1) for best case. Throughout the paper, we use the approach of comparative examples to show higher accuracy of our algorithms compared to the one proposed in [LL+99]. Theoretical analysis, concrete examples and experimental result show that our algorithms can significantly improve the accuracy and time complexity of the calculation of Field Similarity.

[GUS97] D. Guseld. “Algorithms on Strings, Trees and Sequences”, in Computer Science and Computational Biology. CUP, 1997.

[LL+99] Mong Li Lee, Hongjun Lu, Tok Wang Ling and Yee Teng Ko, "Cleansing data for mining and warehousing", In Proceedings of the 10th International Conference on Database and Expert Systems Applications (DEXA99), pages 751-760,August 1999.

[WF74] R. Wagner and M. Fisher, "The String to String Correction Problem”, JACM 21 pages 168-173, 1974.

* Due to historical reason, in this paper, we equalize two terms “string similarity” and “field similarity”

## 1. Introduction

In many applications, it is necessary to determine the string similarity*. Text comparison [SV94,MSU97,CPSV00,ABR00,MS00,KR87,KMR72,GAL85,ME96] now appears in many disciplines such as compression, pattern recognition, computational biology, Web searching and data cleaning[HS95,BD93]. Edit distance[WF74] approach is a classic method to determine Field Similarity. A well known dynamic programming algorithm [GUS97] is used to calculate edit distance with the time complexity O(nm). (for worst case, average case and even best case) Since then, progress has been made in terms of time complexity such as O(n) [Kar93], $\Omega$(nm) [SV96], O(kn) [LV86, MYE86], O(n poly(k)/m+n) [CH98], O(n$\sqrt{m}$) [ABR87], O(n$\sqrt{k}$) [APL00]. However, all these progresses are obtained by relaxing the problem in a number of ways. Hence, when subsequent comparison is made in this paper with respect to time complexity, we still refer to O(nm) [GUS97]. Instead of continuing with improving the edit distance approach, [LL+99] adopted a brand new approach---token-based approach. Its new concept of token-base-----retain the original semantic information, good time complex----O(nm) (for worst, average and best case) and good experimental performance make it a milestone paper in this area. Further study indicates that there is still room for improvement of its Field Similarity algorithm. Our paper is to introduce a package of substring-based new algorithms to determine Field Similarity. Combined together, our new algorithms not only achieve higher accuracy but also gain the time complexity O(knm) (k<0.75) for worst case, O($\beta$*n) where $\beta$<6 for average case and O(1) for best case. Throughout the paper, we use the approach of comparative examples to show higher accuracy of our algorithms compared to the one proposed in [LL+99]. Theoretical analysis, concrete examples and experimental result show that our algorithms can significantly improve the accuracy and time complexity of the calculation of Field Similarity. The rest of the paper is organized as follows. Section 2 gives a background description of the algorithm of calculating Field Similarity presented in [LL+99]. Section 3 proposes our algorithms of calculating Field Similarity and exhaustively compares the new algorithms with the previous one. Section 4 provides experiments to prove the performance improvement with the introduction of the new algorithms.

* Due to historical reason, in this paper, we equalize two terms "string similarity" and "field similarity"

## 2.Preliminary Background

This section gives a brief description of the algorithm to calculate Field Similarity presented in [LL+99].

Let a field X have words $O_{x1}$, $O_{x2}$,….., $O_{xn}$ and the corresponding field Y have words $O_{y1}$, $O_{y2}$,……, $O_{ym}$. Each word $O_{xi}$ ,$1 \le i \le n$ is compared with words $O_{yj}$, $1 \le j \le n$. let $DoS_{x1}$, $DoS_{x2}$,….., $DoS_{xn}$, $DoS_{y1}$, $DoS_{y2}$,….., $DoS_{ym}$ be the maximum of the degree of similarities for words $O_{x1}$, $O_{x2}$,….., $O_{xn}$, $O_{y1}$, $O_{y2}$,……, $O_{ym}$ respectively. Then the Field Similarity for field X and Y

$$SIM_F(X,Y) = \frac{\sum_{i=1}^{n} DoS_{x_i} + \sum_{j=1}^{m} DoS_{y_j}}{n+m} \qquad (1)$$

About the calculation of degree of similarity of words---DoS:

- If two words are exactly the same, the degree of similarity between these two words is 1.
- If there is a total of x characters in the word, then we deduct $\frac{1}{x}$ from the maximum degree of

similarity of 1 for each character that is not found in the other word. For example, if we compare "kit" and "quit", then $DoS_{kit} = 1 - \frac{1}{3} = 0.67$ since the character k in "kit" is not found in "quit" and $DoS_{quit} = 1 - \frac{2}{4} = 0.5$ since the characters q and u in "quit" are not found in "kit".

Exercise: compute the Field Similarity of the filed "address" of record 1and 2 in table 1.

| Record | Name | Address |
|---|---|---|
| 1 | Qi Xiao Yang | 129 Industry Park |
| 2 | Qi Xiao Yang | 129 Indisttry Park |

Table 1 calculation of degree of similarity of words

1. The degree of similarity between "129" and "129" is 1, between "129" and "Indisttry" is 0, between "129" and "Park " is 0. So according to the above rule, $DoS129_{R1}$ =1. (***DoS should be the maximum***)

2. The degree of similarity between "Industry" and "129" is 0, between "Industry" and "Indisttry" is 1-$\frac{1}{8}$=0.875, between "Industry" and "Park " is 0. So according to the above rule, $DoS\text{Industry}_{R1}$ =0.875.

3. In the same way, we will obtain the following:

$$DoS\text{Park}_{R1}=1,\ DoS\text{129}_{R2}=1,\ DoS\text{Indisttry}_{R2}=1-\frac{2}{9}=0.778,\ DoS\text{Park}_{R2}=1$$

4. When Formula 1 is employed, the address Field Similarity for R1 and R2 can be obtained as:

$$SIM_F(X,Y)=\frac{\sum_{i=1}^{n} DoS_{x_i}+\sum_{j=1}^{m} DoS_{y_j}}{n+m}=\frac{1+0.875+1+1+0.778+1}{6}=0.942$$

## 3 Proposed new Field Similarity algorithm

This section proposes a new algorithm----Moving Contracting Window Pattern Algorithm (MCWPA) to calculate Field Similarity. Firstly, we give the definition of **window pattern.** All characters as a whole within the window constitute a **window pattern**. Take a string "abcde" as an example, when the window is sliding from left to right with the window size being 3, the series of window patterns obtained are "abc", "bcd" and "cde".

Let a field X have n characters (including blank space or comma, this applies to the following) and the corresponding field Y have m characters. *w* represents window size, Fx represents the field X and Fy represents the field Y. The Field Similarity for Fx and Fy is

$$SIM_F(X,Y)=\sqrt{\frac{SSNC}{(n+m)^2}} \tag{2}$$

*SSNC* represents the Sum of the Square of the Number of the same Characters between Fx and Fy. $SIM_F(X,Y)$ approximately reflects the ratio of the total number of the common characters in two fields to the total number of characters in two fields.

Imagine we have two windows, one for each field. The basic idea is that we begin with big window size. If window pattern in field 1 is the same as that in field 2, we record the contribution of this matching in SSNC and mark these window patterns as inaccessible to avoid revisiting in the following rounds. Every next round, window size decreases by 1. And within one round, as searching for the same window pattern is going on, windows move from left to right.

The following is the complete algorithm (MCWPA) to calculate *SSNC.*

```
w= the smaller of n and m;
SSNC=0;
Fs=the smaller of Fx and Fy;
window is placed on the leftmost position;
while ((window size is not 0) or (still some characters in Fs are accessible))
{
    while (window right border does not exceed the right border of the Fs )
    {
        if ( the window pattern in Fx has the same pattern anywhere in Fy )
        {
            SSNC= SSNC +(2w) 2 ;
            mark the pattern characters in Fx and Fy as inaccessible characters to avoid revisiting;
        }
        move window rightward by 1 (if the window left border is on an inaccessible character, move window rightward by 2 and so on and so forth)
    }
    w=w-1;
    window is placed on the leftmost position where the window left border is on an accessible character;
}
return SSNC;
```

Figure 1 MCWPA algorithm

The following example is provided to illustrate how to calculate the Field Similarity with MCWPA and formula(2).

**Example 1**: calculate the following Field Similarity.

| | |
|---|---|
| Field 1 | abc de |
| Field 2 | abc k de |

The process of calculating SSNC with MCWPA is shown is figure 2 in detail (next page).

$$SIM_F(X,Y)= \sqrt{\frac{SSNC}{(n+m)^2}} = \sqrt{\frac{(2*4)^2+(2*2)^2}{(6+8)^2}} \approx 63\%$$

**Exercise** what is the Field Similarity between the field1 "abcd" and field2 "abcd"? (the answer is 100%.)

## 3.1 Analysis and Comparison of Two Algorithms of Field Similarity

This section will give some examples to show that MCWPA can overcome some drawbacks that exist in the previous algorithm of the Field Similarity. Also the logic behind the design of MCWPA is presented.

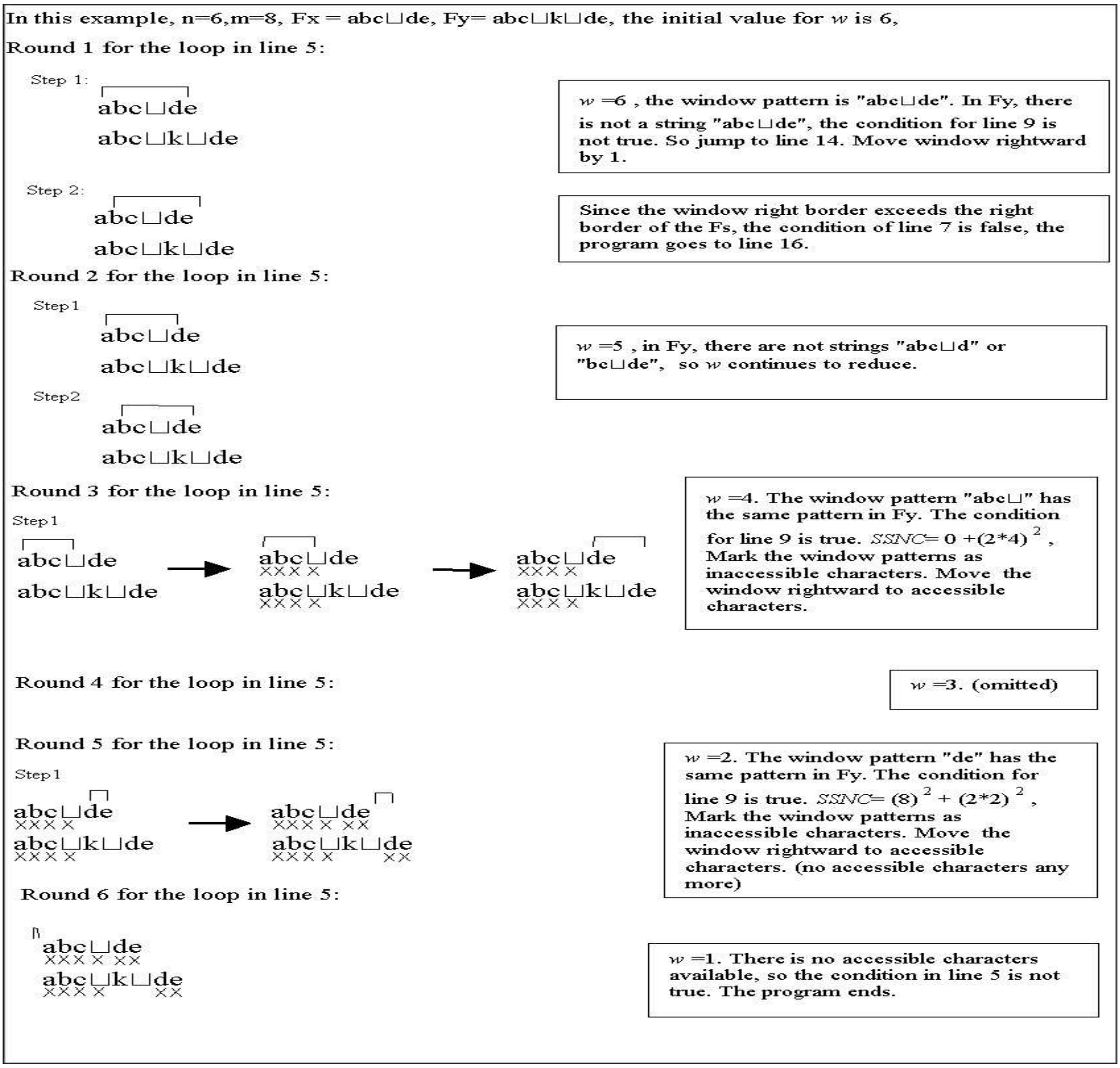

Figure 2 The process of calculating SSNC with MCWPA for example 1

**Example 2**: calculate the following Field Similarity with the above two algorithms.

| | |
|---|---|
| Field 1 | ex ex ex ex ex ex ex ex ex ex |
| Field 2 | ab ab ab ab ab ab ab ab ab ex |

With the previous algorithm,

$$\mathrm{SIM_F(X,Y)} = \frac{\sum_{i=1}^{n} DoS_{x_i} + \sum_{j=1}^{m} DoS_{y_j}}{n+m} = \frac{11}{20} > 50\%$$

With MCWPA,

$$\mathrm{SIM_F(X,Y)} = \sqrt{\frac{SSNC}{(n+m)^2}} = \sqrt{\frac{(2*3)^2}{(2*29)^2}} = \frac{3}{29} \approx 10\%$$

Obviously, the two fields are quite different, only 10% common characters. However, the result of the previous algorithm shows that these two fields have 50% similarity. In contrast, the result of MCWPA is about 10%, which is quite close to the expectation.

**Analysis:** This example shows that there is a drawback for the previous algorithm. In it, $DoS_{x1}$, $DoS_{x2}$,….., $DoS_{xn}$, $DoS_{y1}$, $DoS_{y2}$,….., $DoS_{ym}$ are the **maximum** of the degree of similarities for words $O_{x1}$, $O_{x2}$,….., $O_{xn}$, $O_{y1}$, $O_{y2}$,……, $O_{ym}$ respectively. If quite a number of words in one field are similar to only one word in the other field and dissimilar to other words, the previous algorithm will give inaccurate result. MCWPA overcomes this problem by marking the same characters in two fields as inaccessible so as to avoid revisiting.

**Example 3**: calculate the following Field Similarity for two cases with the above two algorithms.

Case1:

| | |
|---|---|
| Field 1 | de⊔abc |
| Field 2 | de⊔abc |

Case2:

| | |
|---|---|
| Field 1 | abc⊔de |
| Field 2 | de⊔abc |

With the previous algorithm for case 1: $\mathrm{SIM_F(X,Y)} = 1$,

for case2: $\mathrm{SIM_F(X,Y)} = 1$

With MCWPA for case 1:

SIM$_F$(X,Y) = $\sqrt{\frac{SSNC}{(n+m)^2}}$ = $\sqrt{\frac{(2*6)^2}{(6+6)^2}}$ =1,

for case 2:

SIM$_F$(X,Y) = $\sqrt{\frac{SSNC}{(n+m)^2}}$ = $\sqrt{\frac{(2*3)^2+(2*2)^2}{(6+6)^2}}$

=0.6≈60%

**Note:** for case1, two algorithms produce the same result.

**Analysis:** Clearly, the similarity in case 1 should be higher than that in case 2. However, the same results based on the previous algorithm suggest that the previous algorithm considers "abc⊔de" and "de⊔abc" in case 2 the same. This disagrees with our common sense. In the following experiment section, we will show that this is fatally erroneous in some dataset with Chinese names. Further study of the previous algorithm shows that the adoption of word as basic unit results in its inability to distinguish between two exactly the same fields and two fields with the same words in different sequences. To improve the accuracy, MCWPA is based on substring and uses the character as the unit. In this example, if the unit is word, both case 1 and case 2 have two same words. In contrast, if the unit is character, case 1 has 6 same characters and case 2 has 5 same characters. As expected, SIM$_F$(X,Y) in case 1 is larger than SIM$_F$(X,Y) in case 2 when MCWPA is employed.

**Example 4**: calculate the following Field Similarity for two cases with the above two algorithms.

Case1:

| | |
|---|---|
| Field 1 | Fu Hui |
| Field 2 | Mr Fu Hui |

Case2:

| | |
|---|---|
| Field 1 | Fu Hui |
| Field 2 | Fu Mr Hui |

With the previous algorithm for case 1: SIM$_F$(X,Y) =80%,

for case 2: SIM$_F$(X,Y) =80%,

With MCWPA for case 1:

$$\text{SIM}_F(X,Y) = \sqrt{\frac{SSNC}{(n+m)^2}} = \sqrt{\frac{(2*6)^2}{(6+9)^2}} = 80\%,$$

for case 2:

$$\text{SIM}_F(X,Y) = \sqrt{\frac{SSNC}{(n+m)^2}} = \sqrt{\frac{(2*2)^2+(2*4)^2}{(6+9)^2}} \approx 60\%$$

**Note:** for case1, two algorithms produce the same result.

**Analysis:** Intuitively, in case 1, "Fu Hui" and "Mr Fu Hui" should be the same person. In case 2, the likelihood exists that due to transposition error, originally "Fu Mr Hui" should be " Mr Fu Hui". However, in more likelihood, due to typographical errors, originally "Fu Mr Hui" should be " Fu Mi Hui" or "Fu Ma Hui", etc. Factually, the two common words "Fu Hui" in field 2 of case 1 are continuous. In contrast, in field 2 of case 2, they are interpolated by another word "Mr", hence the similarity between two fields is severely reduced. Thus intuitively and factually two fields in case 1 should be more similar than those in case 2. However, the previous algorithm gives the same results for case 1 and case 2. In contrast, the results based on MCWPA show that the similarity for case 1 is reasonably higher than that for case 2. With respect to characters, both case 1 and case 2 have 6 common characters ("Fu" "⊔Hui"). According to example 3, even MCWPA can not distinguish case1 from case 2. Further examination of the two cases reveals that in field 2 of case 1, these 6 characters are continuous while in field 2 of case 2, they are not. In order to reflect the difference in terms of continuity despite the same number of common characters, MCWPA introduces the square to the calculation of $\text{SIM}_F(X,Y)$. In the calculation of $\text{SIM}_F(X,Y)$ in example 4 with MCWPA, the fundamental reason that the result of case1 is larger than that of case2 is because $6^2 > 2^2+4^2$. Mathematically, it is easily seen that the square of the sum of numbers is larger than the sum of the square of numbers, that is, $(a+b+\ldots+n)^2 > a^2+b^2+\ldots\ldots+n^2$, (if $a \neq b \ldots\ldots \neq n \neq 0$). In this way, the introduction of square in the calculation of $\text{SIM}_F(X,Y)$ can overcome the continuity problem which leads to the inaccurate result for the previous algorithm.

### 3.2 The Comparison of Time Complexity between two Algorithms

For pedagogical reasons, suppose we have two fields with the same number of words (W) and same number of characters (N).

**3.2.1 For the previous algorithm:**

Since every word in one field needs to be compared with every word in the other field to find the maximum DoS, the complexity for this step is $O(W^2)$. The complexity of determining whether every character in one word is in the other word is $O((\frac{N}{W})^2)$. Both fields need to be calculated. Therefore, the total time complexity of calculation of Field Similarity by the previous algorithm is $2*O(W^2)*O((\frac{N}{W})^2)$, namely, $2*O(N^2)$, no matter it is worst case, average case or best case.

**3.2.2 For MCWPA:**

Some preparatory knowledge is provided as follows:

When the window size is N, the complexity is $O(1^2)$.

When the window size is N-1, the complexity is $O(2^2)$.

⋮

When the window size is 1, the complexity is $O(N^2)$.

We will discuss the following two situations: 1) with user-specified $SIM_F(X,Y)$ Threshold (ST) 2) without user-specified ST. Since situation 1 is more common and therefore of more practical and theoretical value, it should and does deserve more space in our paper.

**3.2.2.1 With user-specified $SIM_F(X,Y)$ threshold (ST):**

**3.2.2.1.1 UBWS and LBWS**

From figure 1, we know that MCWPA begins with the window size N and carries on with N-1, N-2…… Now, with the knowledge of ST, can we begin directly with a window size named Upper-Bound Window Size (UBWS) so that if there are matching strings (length L) longer than or equal to the UBWS, we can safely determine that the two fields are duplicate. The following presents how to get UBWS with the user-specified ST.

With formula 2,

$$SIM_F(X,Y)=\sqrt{\frac{SSNC}{(n+m)^2}}=\sqrt{\frac{(2*L)^2}{(N+N)^2}}$$

Suppose there exists a USWS that makes

$$ST=\sqrt{\frac{(2*\text{UBWS})^2}{(N+N)^2}} \qquad (3)$$

Since L ≥ UBWS is true, $SIM_F(X,Y) \geq ST$ is also true. This means, with L ≥ UBWS, we can safely determine the two fields are duplicate.

Based on (3), we can get

UBWS= N*ST (4)

**Example 5**: compare the number of comparisons involved in determining whether two fields are duplicate by two algorithms if the $SIM_F(X,Y)$ threshold (ST) is 0.8.

| | |
|---|---|
| Field 1 | abcdefgh ijklmnpo |
| Field 2 | abcdefgh ijklmnwo |

With the previous algorithm:

According to the above analysis, the total number of operations is $2*2^2*(\frac{16}{2})^2$ =512 (2 words for each field).

With revised version of MCWPA:

With formula 4: UBWS =N* ST=17*0.8≈14

As mentioned before, revised MCWPA algorithm skips bigger size window and only uses window size 14 to detect whether there are matching strings. Since the matching strings “abcdefgh ijklmn” are 15-character-long, the algorithm can find the matching strings “abcdefgh ijklm” in the first step and come to the conclusion that these two fields are duplicate. So the total number of comparisons is 1.

What if there are not matching strings longer than UBWS? We need to continue with smaller size windows as described in figure 1. As with the idea of UBWS, can we possibly find a window size named Lower-Bound Window Size (LBWS). With LBWS, even though the field is full of maximum possible matching strings all equally with the length being LBWS and remaining matching strings, the $SIM_F(X,Y)$ still can not meet ST. For example, for two strings A=”abcdefghij” and B=”ghidefabcj”, even though there are three 3-character-long matching strings and one 1-character-long remaining matching string, namely, “abc”, “def”, “ghi” and “j”, the $SIM_F(X,Y)$ between these two fields still can not meet the ST=0.55. ($SIM_F(X,Y)=\sqrt{\frac{(2*3)^2+(2*3)^2+(2*3)^2+(2*1)^2}{(10+10)^2}}$ =0.529) Also, it is easily seen that 3 is the maximum possible length because $SIM_F(X,Y)=\sqrt{\frac{(2*4)^2+(2*4)^2+(2*2)^2}{(10+10)^2}}$ =0.6>0.55=ST. Thus, for ST=0.55, the LBWS is 3. So for this example, we directly use window size 4 to detect whether there are matching strings. If not, we conclude that these two strings are not duplicate. Generally, let the Length of the Remaining matching strings be RL, obviously, RL=N-(N/LBWS)*LBWS.

With formula (2)

$$SIM_F(X,Y)=\sqrt{\frac{SSNC}{(n+m)^2}}=\sqrt{\frac{(2*LBWS)^2+(2*LBWS)^2+.....(2*RL)^2}{(N+N)^2}}=$$

$$\sqrt{\frac{(2*LBWS)^2*(N/LBWS)+(2*(N-(N/LBWS)*LBWS))^2}{(N+N)^2}}$$

$$=\sqrt{\frac{(LBWS)^2*(N/LBWS)+(N-(N/LBWS)*LBWS)^2}{N^2}}\leq ST \quad (5)$$

With formula (5), we can easily get LBWS with given ST by testing every value between 1 and UBWS, the time complexity is only O(UBWS).

**3.2.2.1.2 Expandable Region Match Algorithm (ERMA)**

It can be seen that the core of UBWS, LBWS and MCWPA technology is to find the matching strings efficiently. In this subsection, we propose an algorithm--Expandable Region Match Algorithm (ERMA). It can collect information for all matching strings at O(3N) for best case, $O(k*N^2)$ for worst case (k<75%) and $O(\beta*N)$ ($\beta<6$) for average case. First, we present an introductory example to demonstrate the rough idea of ERMA. How to find the matching substring “ab” with ERMA for field 1)"xxxabxx" and field 2)"yabyyyy"? Suppose now we have already had character information about field 2, that is, “y” is in position 1 of field 2, “a” is in position 2 of field 2……the last character “y” is in position 7 of field 2 and there is no “c”, “d”…. “x” in field2. When we search for matching strings in field 1 character by character, we can easily know that the first three “x”s have no counterparts in field 2. When it comes to the fourth character “a”, we know that we have a character “a” in the second position of field 2. Next, we compare the fifth position of field 1 with the third position of field 2 and we find anther common character “b”. When we compare the sixth position of field 1 with the fourth position of field 2, we find they are not the same. Thus, we find the matching substring “ab”. The crucial point for ERMA is that we must have position information for every character in field 2 in advance. Next, we introduce the ERMA in detail by several examples. For illustrative reasons, both fields consist of only ordinary characters (a—z).

**Example 6**: locate all matching strings by ERMA.

| | |
|---|---|
| Field 1 | akabc axyz mo |
| Field 2 | aabc axyz muo |

**Step 1---pre-process (Regionalize Field 2).**

Imagine we have a character-region with 26 sub-regions, namely, “a” sub-region, “b” sub-region….We start with position 1 , 2, 3….of the field 2 (excluding blank space), put character “x” into x sub-region with the character’s position information. For example 6, the result after step1 is shown in figure 3.

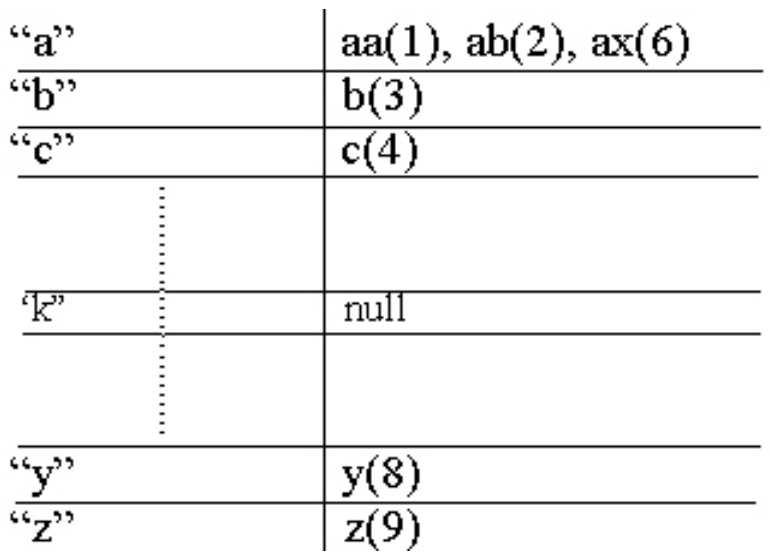

Figure 3 A character-region with Capacity Limit ≥ 3

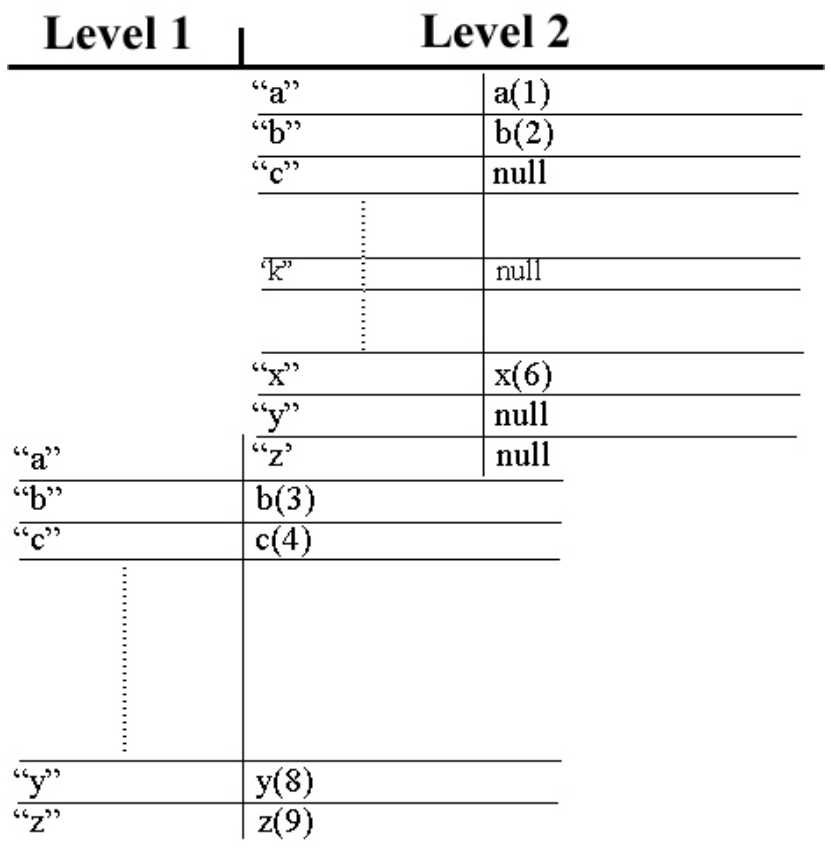

Figure 4 A character-region with Capacity Limit 1

Since “b” is in position 3, b(3) is put into “b” sub-region. In “a” subsection, there are 3 elements---“aa”, ”ab” and ”ax” since there are three “a” occurrences in field 2. Note that ax(6) indicate that the position of “a” is 6 not that of “x”. Capacity Limit for a character-region is the upper limit of the number of elements for the subregion. If the Capacity Limit for Figure 3 is 1, we need to further partition “a” subregion----expand “a” subregion. The result after expansion is shown in Figure 4.

**Step 2---process every character in field 1.**

In particular, for every character in field 1: 1)get the longest matching strings starting from that character based on the character-region built in step1. 2)Record the information of length of longest matching strings starting from that character and the corresponding starting position in field 2 . For example 6, we begin with the string starting with the first character “a”, namely “akabc….”. Based on the character-region shown in Figure 4, the first character “a” has 3 common characters, while the second character “k” meets with a “null” in the level 2 of character-region. This means that the string starting with the first character “a” only has 1-character-long longest matching string. Since the longest matching string “a” has 3 occurrences in field 2, we randomly choose any one of them. The reason why we randomly choose is given in section 3.2.2.1.3. In practice, to guarantee that they can be chosen with equal probability, machine generated random numbers with equal probability are used to make the decision. A record is then generated with information that the length is 1 and the position is any one of the three choices “1”, “2” and “6”, say, “2”. And this record is linked to the first character “a”. (see figure 5) Easily seen, the string starting with the second character “k” does not have any matching string. For the string starting with the third character “a”, namely, “abc axyz mo”, similarly, based on the character-region shown in Figure 4, the first character “a” has 3 common characters, while the second character “b” meets with a “b” in the level 2 of character-region with a pointer pointing to position 2 of field 2. Base on this information, next, the string “abc axyz m…” in field 1 compares with the string “abc axyz m…” which starts from position 2 of field 2. This comparison results in a 10-character-long matching string “abc axyz m”. Similarly, A record is generated with information that the length is 10 and the position is 2. This record represents that a substring starting with “a” in field 1 has a 10-character-long matching string in field 2 that starts from position 2 of field 2. And this record is linked to the third character “a”. Since we have processed the first character of the 10-character-long matching string, we can skip comparisons for the next 9 consecutive characters (“bc axyz m”) by the following technique----**Expectation**. If a character is in its expected position, we don’t need to make comparison for it. Take the fourth character “b” as an example. The expected position for it is 3 in field 2 since “b” belongs to an existing matching string which starts from “a” and the position of “a” is 2 and “b” is next to it. Based on the character-region shown in Figure 4, we find that the character “b” only has one occurrence----position 3 which is the same as its expected position, obviously, we can skip comparison for it. If there are several occurrences of “b”-----this phenomenon is called **Conflict Type 1---** although we can skip comparison for “b” which is in expected position, for other occurrences of “b”, the approach of processing them is the same as that of processing the third character “a”. The result after step 2 is shown in figure 5. The top level is a group of pairs representing the length of longest matching strings starting with that particular character and the corresponding position in field 2.

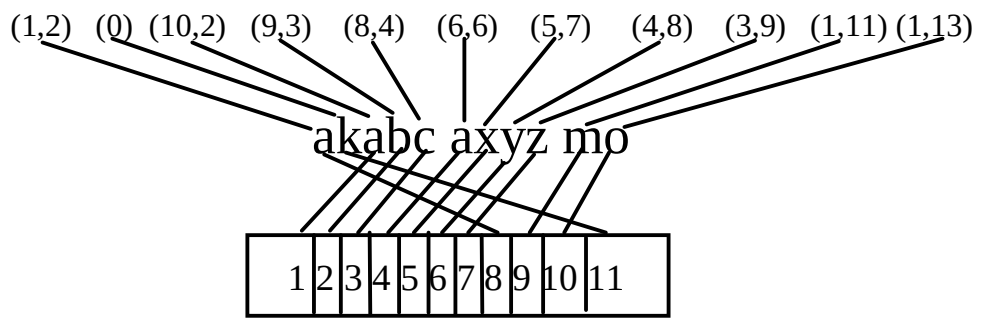

Figure 5 The result after step 2

**Step 3---post-process (find all matching strings for the current round)**

Based on the information from step 2, we can easily get the process sequence by sorting characters into descending order according to the length of longest matching strings starting from that particular character. For example, In figure 5, since the length value (10) of the third character “a” is largest, this character should be processed first. The sorting result is shown as a train of numbers on the bottom of figure 5 that indicate the process sequence. The process starts with the character “a” since the first number in the train points to it. On the other hand, this character “a” is also the starting character of the longest matching strings for this processing round. We mark 10 consecutive characters in two fields starting from “a” as inaccessible. Correspondingly, all numbers in the train that link from these inaccessible characters are also marked as inaccessible. The result is shown in figure 6.

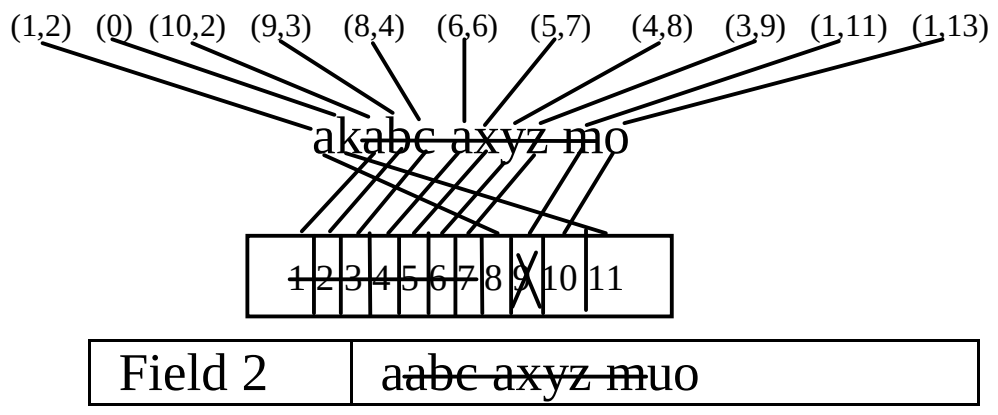

Figure 6 The result after 10 characters are marked in step 3

We continue the current round with the leftmost accessible number in the train. For figure 6, it is “8” which points to “a”. The information on the top level of figure 6 about the character “a” indicates that this character has a matching string starting from position 2 of field 2. Unfortunately, the character in this position has been marked as inaccessible, which means this character has already belonged to another matching string. This phenomenon is called **Conflict Type 2**. The solution to Conflict Type 2 is that if we find that a character “x” with length “l” has been marked as inaccessible, we ignore processing “x” and continue to process other characters with the same length “l”. After all characters with length “l” are finished, we go to a new round by repeating step 2 and step 3, but all inaccessible characters are not processed any more. In figure 6, we continue with the next accessible number “10” in the train. It points to “o” and the length of “o” is also 1, so we find another matching strings and mark them in two fields. Since the length of the character “k” linked from the next accessible number in the train is 0 and less than 1, the current round ends.

**3.2.2.1.3 Implementation of ERMA and Time Complexity**

For step 1, there are two types of implementation: 1) Fixed size (26) array to represent character-region with Capacity Limit equal to 1. 2) A tree whose nodes have no more than 26 children. The disadvantage for the array-based implementation is more storage. For example, in Figure 3, it needs to store “k” even though k’s value is “null” while tree-based implementation does not. The advantage coupled with the space disadvantage is faster search. For example, to find “c”, we simply check whether array [3] is “null” or not because “c” is the 3rd alphabetically. While for the tree-based implementation, along the path to find the leaf, comparison needs to be made at non-leaf nodes even though it is negligibly cheap. The character-region with either of these two types of data structures can be built in O(N) time. In addition, another choice is Fixed size (26) array to represent character-region with Capacity Limit greater than 1. It is a compromise between array implementation and tree implementation with regard to time and space.

For step 2, if there is no conflict type 1, we can collect information for all characters in field 1 at O(N). In worst case where there is heavy conflict, the time complexity is $O(k*N^2)$. ($k<50\%$) (for example, field 1 is “abababab” and field 2 is “aaaaaaaa”) In average case, empirically and experimentally, the conflict type 1 occurs within small scope, so the time complexity is $O(\beta*N)$ where $\beta<2$.

For step 3, when we sort characters according to the length of longest matching strings starting from that particular character, we can use Radix sort approach[CP01]. The time complexity for Radix sort is O(N). If there is no conflict type 2, one round is enough to find all matching strings. The time complexity is O(1). In worst case where there is heavy conflict, because the number is randomly chosen as mentioned before, the time complexity is $O(k*N^2)$. (Empirically and experimentally, $k<25\%$) (for example, field 1 is “abababab” and field 2 is “aaaaaaaa”) In average case, we can find all matching strings within 2 rounds. The time complexity is correspondingly for step3 O( $\beta$ *N) where $\beta$ <2.

**3.2.2.1.4 Summary of the Situation with Given ST**

Having introduced the definition of $SIM_F(X,Y)$, the method of calculating $SIM_F(X,Y)$ with MCWPA, the concepts of UBWS, LBWS and ERMA, we sum up the discussion of the situation where ST is specified as follows. (see figure 7)

Generally, We have two choices, MCWPA and ERMA. To determine whether there are matching strings at least equal to UBWS, MCWPA needs at least $UBWS*(N-UBWS+1)^2$, while in average case, ERMA needs 6N. We choose the smaller of $UBWS*(N-UBWS+1)^2$ and 6N as our scheme.

- For MCWPA, if we find matching strings at least equal to UBWS, we conclude that two fields are duplicate. If we can not find, we need to make choice once again. One is continue with MCWPA with $LBWS*(N-LBWS)^2$. The other is ERMA with 6N. We choose the smaller of $LBWS*(N-LBWS)^2$ and 6N as our scheme. If MCWPA is our choice, we use window size equal to LBWS to search for matching strings. If we can not find, we conclude that two fields are NOT duplicate. If unfortunately we can find, the situation will be quite complicated, we switch to ERMA.
- For ERMA, (we discuss average case in terms of conflict type 1) (1) if there is not conflict type 2, without going to the next round, we can come to the conclusion. After we finish the first round with O(5N), we compare the $SIM_F(X,Y)$ resulting from the contribution of all matching strings from the first round with ST. If it is greater than ST, we conclude that two fields are duplicate. If it is not, we conclude that two fields are NOT duplicate. (2) if there is conflict type 2 , after we finish the first round with O(5N), we compare the $SIM_F(X,Y)$ resulting from the contribution of all matching strings from the first round with ST. If it is greater than ST, we conclude that two fields are duplicate. If it is not, we compare the longest matching string from the first round with LBWS. If it is shorter than LBWS, we conclude that two fields are NOT duplicate. If it is no shorter than LBWS, we must go on to the second round. Taking the contribution of all matching strings from the first round into account, with formula 2 and formula 5, we can get new LBWS (see example 7). The discussion of situation where there is not conflict type 2 is the same as before. We omit it since it is straightforward. If there is still conflict type 2, after we finish the second round with O(N), (because we can use the character-region derived from the first round and we do not need to process inaccessible characters) we compare the $SIM_F(X,Y)$ which is the sum of the contribution of all matching strings from the second round and the first round with the ST. If it is greater than the ST, we conclude that two fields are duplicate. If it is not, we compare the longest matching string from the second round with the new LBWS. If it is shorter than the new LBWS, we conclude that two fields are NOT duplicate. If it is not shorter than the new LBWS, we must go on to the third round. The same process will carry on until either we can come to the conclusion whether they are duplicate or all characters are marked inaccessible. Every more round will cost less and less because more and more characters are marked inaccessible. As discussed before, the time complexity for the worst case is $O(0.25*N^2)+O(0.5*N^2)$ In average case, the time complexity is $O(\beta*N)$ where $\beta<6$.

The above discussion is visually presented in figure 7 which more clearly shows the following conclusions:

1) Only if $UBWS*(N-UBWS+1)^2<6N$, can MCWPA be used. Hence, MCWPA applies to the situation where ST is quite high and the number of comparisons is quite small. The best case O(1) is obtained from MCWPA.
2) For ERMA (right-lower area), if there is not conflict type 2, we can safely reach the conclusion with <5N.
3) For ERMA, if there is conflict type 2 and we come to the conclusion within the first round, the time complexity is <5N. If we come to the conclusion within the second round, the time complexity is <5N+N. Empirically, in average the whole process will end within 3 rounds which corresponds to about 6N.

**Example 7**: calculate the complexity of judging whether the following two fields are duplicate, given that ST is 0.48. (for clarity, we mark the matching strings in two fields)

| | |
|---|---|
| Field 1 | abcdefagha |
| Field 2 | aijklamabc |

**Answer:** Since there are 10 characters, N=10. Based on formula 4, we have UBWS=5. Bases on formula 5, LBWS=2. Because $UBWS*(N-UBWS+1)^2=180>6*N=60$, we choose ERMA instead of MCWPA. Suppose unfortunately, due to conflict type 2, in the first round, we only find the match string "abc".

$SIM_F(X,Y)=0.3<ST=0.48$ and MMSLFC=3>LBWSfC=2. (see figure 7) Thus, we must go to the second round where we find another matching string "a". Once again, unluckily, suppose we encounter conflict. $SIM_F(X,Y)$ of SCAMSUC is $\sqrt{\frac{(2*3)^2+(2*1)^2}{(10+10)^2}}=0.316<ST$, so we need to judge whether MMSLFC<LBWSfC. Since the only matching string is "a", MMSLFC is 1. For LBWSfC, according to the above discussion, we need to try the character length 2 and 1, since LBWS for the first round is already 2. First we try 2 with formula 5. ($SCAPR^*$ represents Sum of Contribution from All Past Rounds, it is equal to $(2*3)^2$ in this case )

$$SIM_F(X,Y)=\sqrt{\frac{SCAPR^*+(2*LBWS)^2.......+(2*RL)^2}{(10+10)^2}}=$$

$$\sqrt{\frac{(2*3)^2+(2*2)^2+(2*2)^2+(2*2)^2+(2*1)^2}{(10+10)^2}}=0.469<ST,$$

so the new LBWS is 2, MMSLFC=1<LBWSfC=2 and we can come to the conclusion that the two fields are unduplicated.

In summary, we reach the conclusion within 2 rounds, the time complexity is 6N=60. In this example, the ST=0.48 is quite low, so MCWPA can not be used. Empirically, if ST is greater than 90%, in majority of the cases, MCWPA will be used. That means, the time complexity will be less than O(6N).

**3.2.2.2 Without User-Specified ST:**

In this situation, because of unavailability of ST, all matching strings need to be found so that formula 2 can be used to calculate $SIM_F(X,Y)$. ERMA is employed to perform this task. Hence part of the above conclusion applies to here. If there is no conflict type 2 (we discuss average case for conflict type 1), within one round, we can find all matching strings. The time complexity for this is O(5N). In worst case where there is heavy conflict type 2, the time complexity is $O(k*N^2)$. (Empirically and experimentally, k<75%) In average case, the time complexity is $O(\beta*N)$ where $\beta<6$.

## 4 Experiment Result

We conducted four sets of experiments with both algorithms. The first dataset is a merger of two datasets that come from two campus surveys conducted through an electronic form within a mass-sent email. The dataset has 782 records. The second dataset is from the 1990 US Census which is a free downloaded dataset coming from http://www.cs.toronto.edu/~delve/data/census-house/desc.html. It has 22784 records. The third and fourth datasets are generated synthetic datasets both with more than 200,000 records. We compare two algorithms by two criteria: 1) Miss Detection (duplicate records are not detected) and 2)False Detection (similar non-duplicate records are treated as duplicate records). The results are presented in figure 8 ~11.

**Analysis:** Experimental results on four datasets consistently indicate that with regard to Miss Detection, the two algorithms perform roughly the same. However, in terms of False Detection, MCWPA performs much better than the previous algorithm. Further study of the testing datasets shows that in the name field, there are some similar non-duplicate names such as "Gao Hua Ming" and "Gao Ming Hua". As analyzed in the example 3, the previous algorithm treats two fields with the same words in different sequences as the matching fields. Thus the high False Detection rate for the previous algorithm begins to make sense. In addition, there are also some similar cases that the previous algorithm treats some names such as "zeng hong" and "zeng zeng" the same. As analyzed in the example 2, MCWPA identifies a large difference in the calculation of Field Similarity between this type of two fields. Generally, from several examples we presented above, the previous algorithm tends to over-evaluate the $SIM_F(X,Y)$, while MCWPA does not. We observe from both experiments that MCWPA is roughly equally effective across the entire range of $SIM_F(X,Y)$ threshold. As opposed to this, the False Detection rate based on the previous algorithm increases significantly as the $SIM_F(X,Y)$ threshold becomes lower and lower. The Miss Detection diagrams show that both algorithms can only perform well in the low $SIM_F(X,Y)$ threshold region. However, the False Detection diagrams indicate that in the low $SIM_F(X,Y)$ threshold region, the False Detection rate from the previous algorithm is very high. This means, with the previous algorithm, if we choose low $SIM_F(X,Y)$ threshold to satisfy Miss Detection rate requirement, we will inevitably obtain poor False Detection performance. This conflict does not show itself in MCWPA.

## Conclusion

This paper has presented a new algorithm (MCWPM) for the calculation of Field Similarity. In essence, MCWPM improves the previous algorithm in the following aspects:

1) The introduction of marking the common characters as inaccessible to avoid revisiting, which is presented in example 2.

2) The adoption of the character as unit for the calculation of Field Similarity instead of words to improve accuracy, which is presented in example 3.

3) The introduction of square to the calculation of Field Similarity to reflect the difference in terms of continuity despite the same number of common characters, which is presented in example 4.

4) The introduction of UBWS, LBWS and ERMA to achieve higher efficiency, which is presented in example 5.

Theoretical analysis, concrete examples and experimental result lead to the conclusion that our new algorithm (MCWPM) can significantly improve the accuracy and efficiency of the calculation of Field Similarity.

## Reference

[ABR00] S. Alstrup, G. S. Brodal and T. Rauhe, “Pattern matching in dynamic texts”, In Proceedings of the 11th Annual Symposium on Discrete Algorithms, pages 819-828, 2000.

[ABR87] K. Abrahamson, “Generalized string matching”, SIAM Journal on Computing, 16(6):1039-1051, 1987.

[ALP00] A. Amir, M. Lewenstein, and E. Porat, “Faster algorithms for string matching with k-mismatches”, In Proceedings of the 11th Annual Symposium on Discrete Algorithms, pages 794-803, 2000.

[BD93] D. Bitton and D.J. DeWitt. “Duplicate record elimination in large data files”, ACM Transactions on Database Systems, 1995.

[CH98] R. Cole and R. Hariharan, “Approximate string matching: A simpler faster algorithm”, In Proceedings of the 9th Annual Symposium on Discrete Algorithms, pages 463-472, 1998.

[CP01] F. M. Carrano and J. J. Prichard, “Data Abstraction and Problem Solving with Java Walls and Mirrors”,2001.

[CPSV00] G. Cormode, M. Paterson, S. C. Sahinalp and U. Vishkin, “Communication complexity of document exchange” In Proceedings of the 11th Symposium on Discrete Algorithms, pages 197-206, 2000.

[Gal85] Z. Galil, “Open problems in stringology”, In Combinatorial Algorithms on Words, pages 1-8. Springer, 1985.

[GUS97] D. Guseld, “Algorithms on Strings, Trees and Sequences”, in Computer Science and Computational Biology. CUP, 1997.

[HS95] M. Hernandez and S. Stolfo, “The merge/purge problem for large databases”, Proc. of ACM SIGMOD Int. Conference on Management of Data pages 127-138, 1995.

[KAR93] H. Karloff, “Fast algorithms for approximately counting mismatches”, Information Processing Letters, 48(2):53-60, November 1993.

[KMR72] R. M. Karp, R. E. Miller, and A. L. Rosenberg, “Rapid identification of repeated patterns in strings, trees and arrays”, In 4th Symposium on Theory of Computing, pages 125-136, 1972.

[KR87] R. M. Karp and M. O. Rabin, “Efficient randomized pattern-matching algorithms”, IBM Journal of Research and Development, 31(2):249-260, March 1987.

[LL+99] Mong Li Lee, Hongjun Lu, Tok Wang Ling and Yee Teng Ko, "Cleansing data for mining and warehousing", In Proceedings of the 10th International Conference on Database and Expert Systems Applications (DEXA99), pages 751-760,August 1999.

[LV86] G. M. Landau and U. Vishkin, “Introducing Efficient Parallelism into Approximate String Matching and a new Serial Algorithm”, In 18th Symposium on Theory of Computing, pages 220-230, 1986.

[ME96] A.E. Monge and C.P. Elkan, “The field matching problem: Algorithms and applications”, Proc. of the 2nd Int. Conference on Knowledge Discovery and Data Mining pages 267-270, 1996.

[MS00] S. Muthukrishnan and S. C. Sahinalp, “Approximate nearest neighbors and sequence comparison with block operations”, In 32nd Symposium on Theory of Computing, 2000.

[MSU97] K. Mehlhorn, R. Sundar and C. Uhrig, “Maintaining dynamic sequences under equality tests in polylogarithmic time”, Algorithmica, 17(2):183-198, February 1997.

[MYE86] E. W. Myers, “An O(ND) difference algorithm and its variations”, Algorithmica, 1:251-256, 1986.

[SV94] S. C. Sahinalp and U. Vishkin, “Symmetry breaking for suffix tree construction”. In 26th Symposium on Theory of Computing, pages 300-309, 1994.

[SV96] S. C. Sahinalp and U. Vishkin, “Efficient approximate and dynamic matching of patterns using a labelling paradigm”. In 37th Symposium on Foundations of Computer Science, pages 320-328, 1996.

[WF74] R. Wagner and M. Fisher, "The String to String Correction Problem”, JACM 21 pages 168-173, 1974.

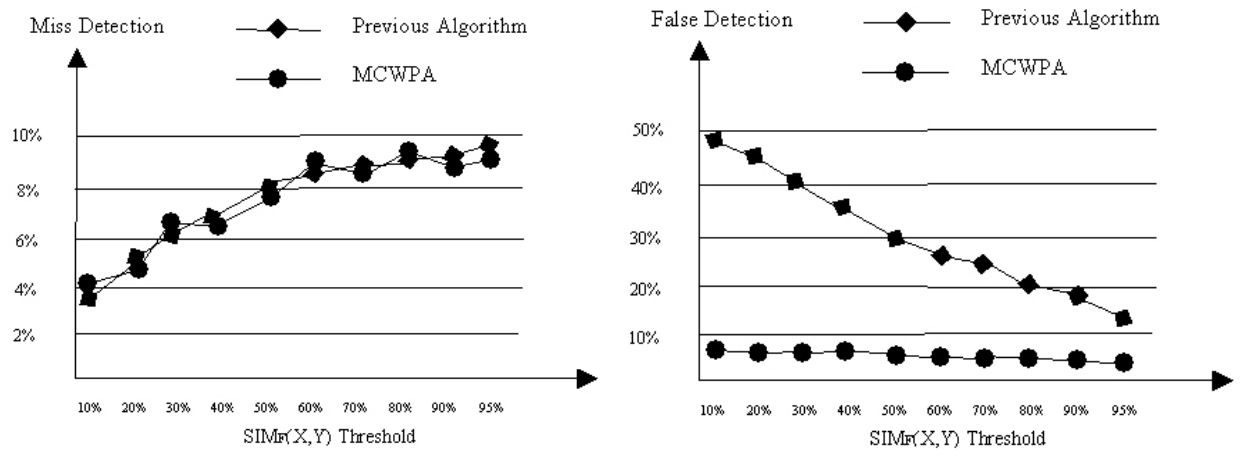

Figure 8 experiment on campus survey dataset

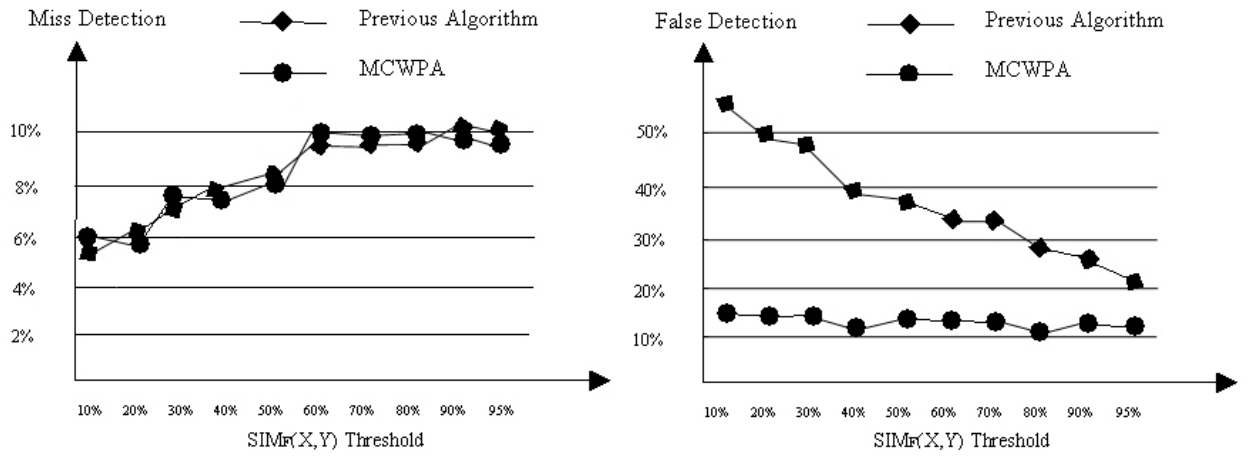

Figure 10 experiment on synthetic dataset 1

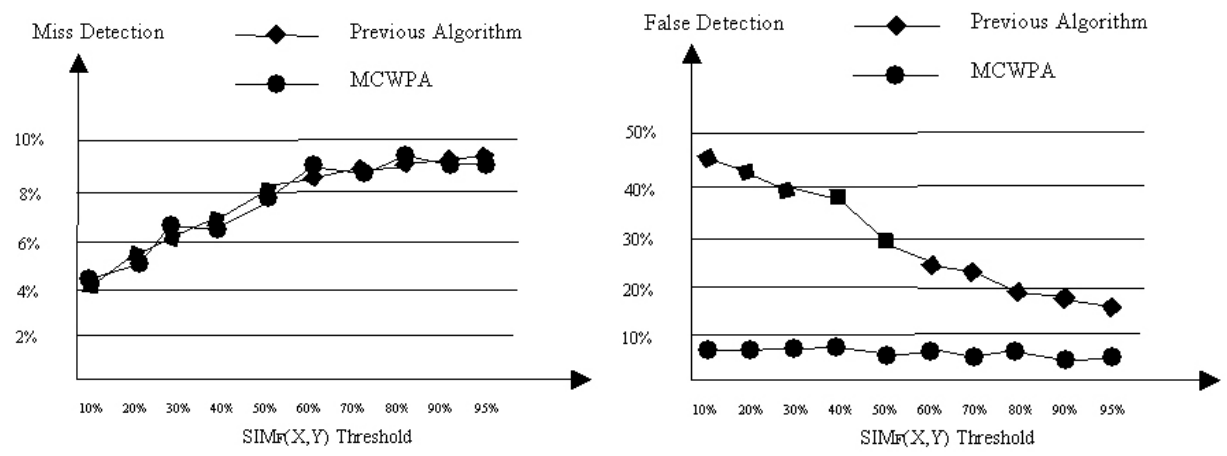

Figure 9 experiment on census dataset

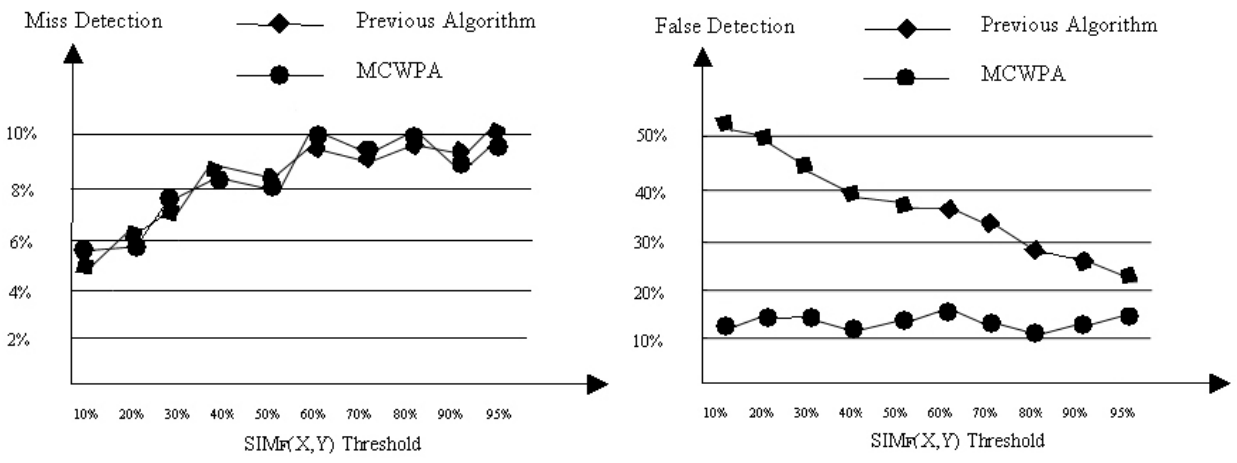

Figure 11 experiment on synthetic dataset 2

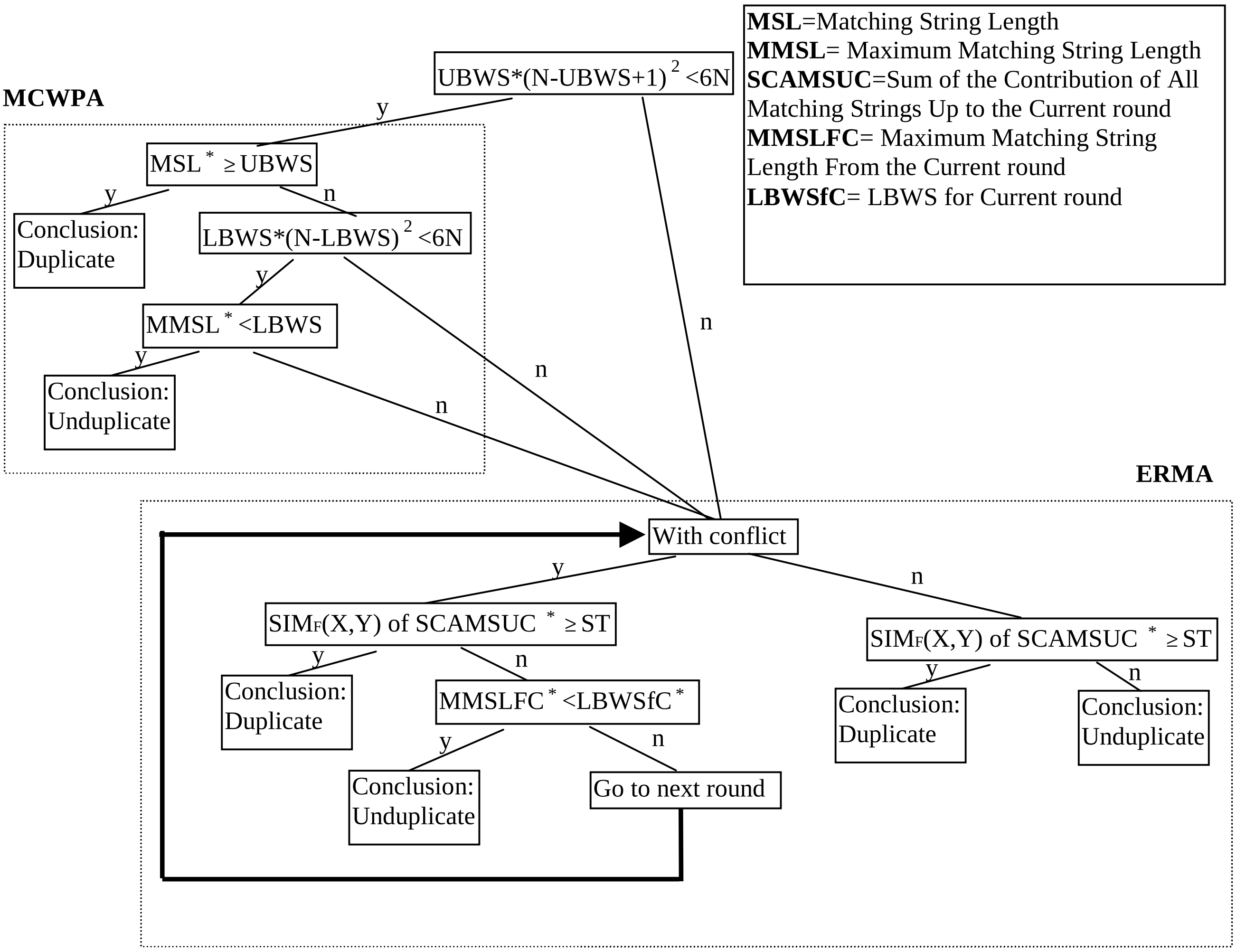

Figure 7 Summary of the discussion of the situation where ST is specified